\documentclass[fleqn,twoside]{article}
\usepackage{espcrc2}
\usepackage{amssymb}
\usepackage[dvips]{graphicx}
\usepackage{psfrag}

\title{Static Potential and Local Color Fields in Unquenched
Three-Dimensional Lattice QCD}

\author{Howard D. Trottier
	\address[MCSD]{Physics Department, Simon Fraser University,
	Burnaby, B.C., Canada, V5A 1S6.} and
        Kit Yan Wong\addressmark}
       
\begin{document}

\begin{abstract}
String breaking by dynamical quarks in (2+1)-d
lattice QCD is demonstrated in this project, by
measuring the static potential and the local color-electric field strength
between a heavy quark and antiquark pair at large separations. Simulations
are done for unquenched SU(2) color with two flavors of staggered quarks.
An improved gluon action is used which allows simulations to be done on
coarse lattices, providing an extremely efficient means to access the quark
separations and propagation times at which string breaking occurs. The static
quark potential is extracted using only Wilson loop operators and hence no
valence quarks are present in the trial states. Results give unambiguous
evidence for string breaking as the static quark potential completely
saturates at twice the heavy-light meson mass at large separations. It is
also shown that the local color-electric field strength between the quark
pair tends toward vacuum values at large separations. Implications of these
results for unquenched simulations of QCD in 4-d are drawn.
\vspace{1pc}
\end{abstract}

\maketitle

In unquenched simulations, we expect the linear rising
static quark potential to saturate at large quark separations $R$.
This is the phenomenon of hadronic string breaking.
However, despite extensive simulations of full QCD by several large
scale collaborations \cite{Aoki98}, a convincing demonstration of
string breaking remains controversial.
Traditionally, Wilson loops have been used to extract the
static quark potential, though recently it has been suggested that this
operator is not suitable because it has a small overlap with
the two-meson ground state at large $R$ \cite{Drum98}.

On the other hand, the Wilson loop produces a heavy quark-antiquark
trial state without explicit light valence quarks. This trial state is of
great physical interest because of the analogy with the process of
hadronization, where light quarks are not present in the initial state,
but rather are materialized from the vacuum.

It was proposed in Ref. \cite{Trot98} that string breaking can in fact be 
observed in Wilson loops by working on coarse lattices using improved actions.
On coarse lattices the computational effort can go towards generating much 
higher statistics, which allows much longer length and time scales
to be accessed. This approach was applied in Ref. \cite{Trot98} to lattice
QCD in (2+1) dimensions, and later to 4-d QCD in Ref. \cite{Dunc93}. It now
appears that the overlap of the Wilson loop with the ground state is large
enough, in the relevant range of $R$, to allow string breaking to be resolved.

The present work is a follow-up on the earlier study of Ref.
\cite{Trot98}. What is new here is a much more extensive study on
larger lattices and with far higher statistics. The results give a much
more convincing demonstration of an asymptote in the static quark
potential at large quark separations. The demonstration of string breaking
from measurements of the local color-electric field strength in unquenched
QCD is entirely new to this work.

To reduce computational cost, we work in 3-d unquenched QCD with
SU(2) color. $\mathrm{QCD_{3}}$ has been shown to share most of the
fundamental properties of the realistic 4-d QCD \cite{Tepe99} and thus
provides excellent insights to QCD.
The coupling $g_{0}^{2}$ in $\mathrm{QCD_{3}}$ has
dimension of mass and so explicitly sets the mass scale for the theory,
i.e., $m\propto g_{0}^{2}$ for any mass quantity $m$. In order to have some
intuition for this mass scale, we express this scale in terms of the length
scale in 4-d --- ``fermi'' ($fm$). Motivated by the fact that
both 3-d and 4-d QCD are linearly confining,
we match the quenched continuum extrapolation of the 3-d string tension,
$\sqrt{\sigma}/g^{2}=0.3353(18)$ \cite{Tepe99}, with the physical value
of the string tension in four dimensions, $\sqrt{\sigma}=0.44GeV$.
This in effect sets the value of $g_{0}$ for our 3-d theory.
A similar value for the coupling is obtained from other 
physical quantities, such as equating the unquenched $\rho$ meson
mass in this theory with the physical value in 4-d QCD.

The tree-level $O(a^{2})$-accurate improved gluon action in 3-d
has the usual form
\begin{flushleft}
\begin{equation}
S_{imp}= -\beta \sum_{x,\mu>\nu} \xi
\left[ \frac{5}{3}P_{\mu\nu}
- \frac{1}{12}\left(R_{\mu\nu}+R_{\nu\mu}\right) \right],
\end{equation}
\end{flushleft}
where $P_{\mu\nu}$ is the plaquette and $R_{\mu\nu}$ is the
2 $\times$ 1 rectangle. The bare lattice anisotropy is entered through
$\xi_{3i}=\xi_{i3}=a_{s}/a_{t}$ and $\xi_{ij}=a_{t}/a_{s}$ ($i,j=1,2$).
We use the staggered quark action in 3-d, which has the same
form as the 4-d action and describes two flavors of four-component
spinors:
\begin{flushleft}
{\setlength\arraycolsep{2pt}
\begin{eqnarray}
S_{KS} = \sum_{x,\mu} \zeta_{\mu} \eta_{\mu}(x) \bar{\chi}(x)
\left[ U_{\mu}(x)\chi(x+\hat{\mu}) \right. \nonumber \\
\left. - U_{\mu}^{\dag}(x-\hat{\mu})\chi(x-\hat{\mu}) \right]
+ 2am_{0} \sum_{x} \bar{\chi}(x)\chi(x),
\end{eqnarray}}
\end{flushleft}
where $\zeta_{3}=a_{s}/a_{t}$ and $\zeta_{1,2}=1$. 

For comparison, most of the simulation parameters employed in the present
project are taken from Ref. \cite{Trot98}. Simulations were done on a
$22^{2}\times 28$ lattice with $\beta=3.0$.
Using the scale setting procedure discussed earlier,
the lattice spacing is identified to be $a_{s}\simeq 0.2fm$ in physical
units. The input bare quark mass in lattice units is scaled according to
$m_{0}/g_{0}^{2}=0.10$, with $m_{\pi}/m_{\rho}$ found to be $\simeq 0.75$.
The input bare anisotropy is $a_{t}/a_{s}=1/2$. The configurations were
generated using the HMD algorithm (the
$\Phi$-algorithm). Overall 30,000 measurements were taken,
corresponding to a total of 300,000 trajectories.

\begin{figure}[!t]
\psfrag{xlabel}[ct][ct]{$R/a_{s}$}
\psfrag{ylabel}[cb][cb]{$a_{s}V(R,T)$}
\psfrag{quenched}[rt][rt]{\tiny quenched}
\psfrag{T=2}[rb][rb]{\tiny $T/a_{s}=2$}
\psfrag{T=6}[rb][rb]{\tiny $T/a_{s}=6$}
\psfrag{T=12}[rb][rb]{\tiny $T/a_{s}=12$}
\psfrag{heavy}[lb][lb]{\tiny heavy-light}
\psfrag{meson}[lb][lb]{\tiny meson mass}
\includegraphics[width=17.5pc]{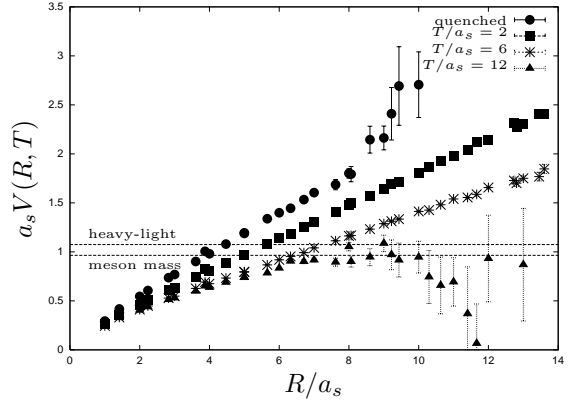}
\caption{The static quark potential $V(R,T)$.}
\label{potential}
\end{figure}

Only (fuzzy) Wilson loops were used to extract the static quark potential.
The final results are presented in Fig. \ref{potential} for
propagation times $T/a_{s}=2, 6, 12$
(or $T\simeq 0.2, 0.6, 1.2fm$ in physical units). The unquenched
heavy-light meson mass was also computed and is shown as
the dotted lines in Fig. \ref{potential}.
The results here give clear indication of string breaking as the unquenched
potential is substantially flattened at large $R$ and
approaches the expected asymptotic value of the heavy-light meson mass.

\begin{figure}[!b]
\psfrag{xlabel}[ct][ct]{$T$}
\psfrag{ylabel}[cb][cb]{$a_{s}^{2}\sigma$}
\includegraphics[width=17.5pc]{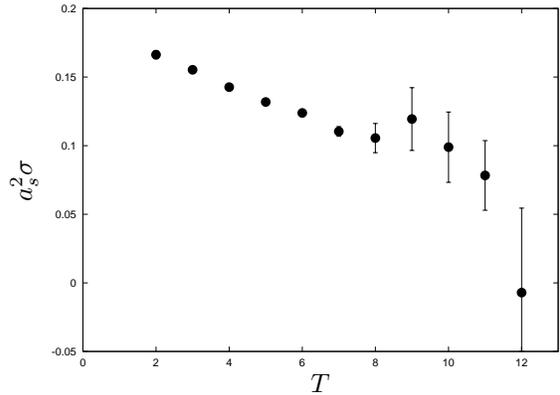}
\caption{String tension $a_{s}^{2}\sigma$ as a function of $T$.}
\label{stringtension}
\end{figure}

To emphasize the flattening of the potential at large values of $R$, we
fitted the data points within $R/a_{s}=8.0-10.0$ with a linear
function $V(R) = \sigma R +b$ and plotted the slope $\sigma$ (string tension)
against the propagation time $T$ in Fig. \ref{stringtension}.
Results here unambiguously show that the slope tends toward zero
at sufficiently large $T$, $T\simeq1.2fm$.

\begin{figure}[!t]
\begin{minipage}[b]{8.5pc}
\psfrag{whatR}[lb][lb]{\tiny $R/a_{s}=5$}
\psfrag{xlabel}[ct][ct]{\tiny $x_{\parallel}/a_{s}$}
\psfrag{ylabel}[cb][cb]{\tiny $E^{\parallel}/E_{max}^{\parallel}$}
\includegraphics[width=8pc]{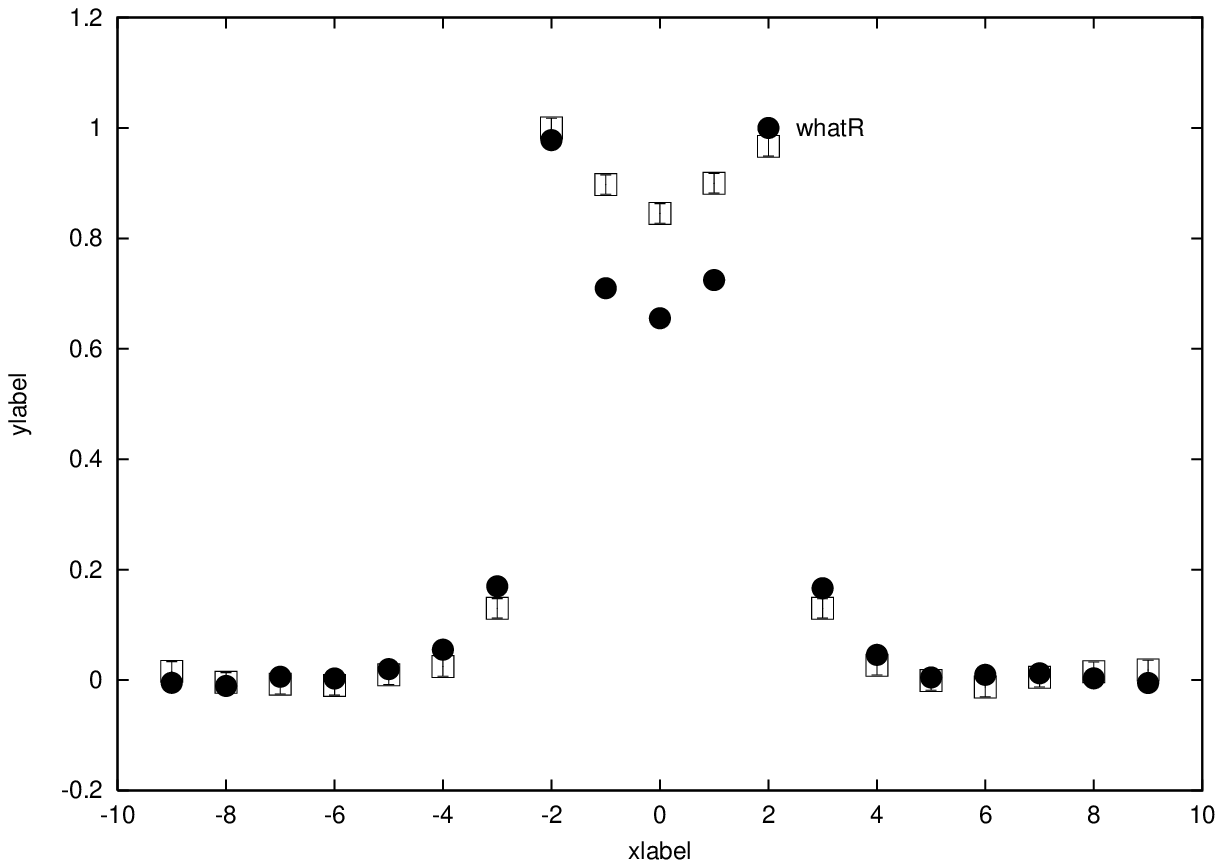}
\end{minipage}%
\begin{minipage}[b]{8.5pc}
\psfrag{whatR}[lb][lb]{\tiny $R/a_{s}=7$}
\psfrag{xlabel}[ct][ct]{\tiny $x_{\parallel}/a_{s}$}
\psfrag{ylabel}[cb][cb]{\tiny $E^{\parallel}/E_{max}^{\parallel}$}
\includegraphics[width=8pc]{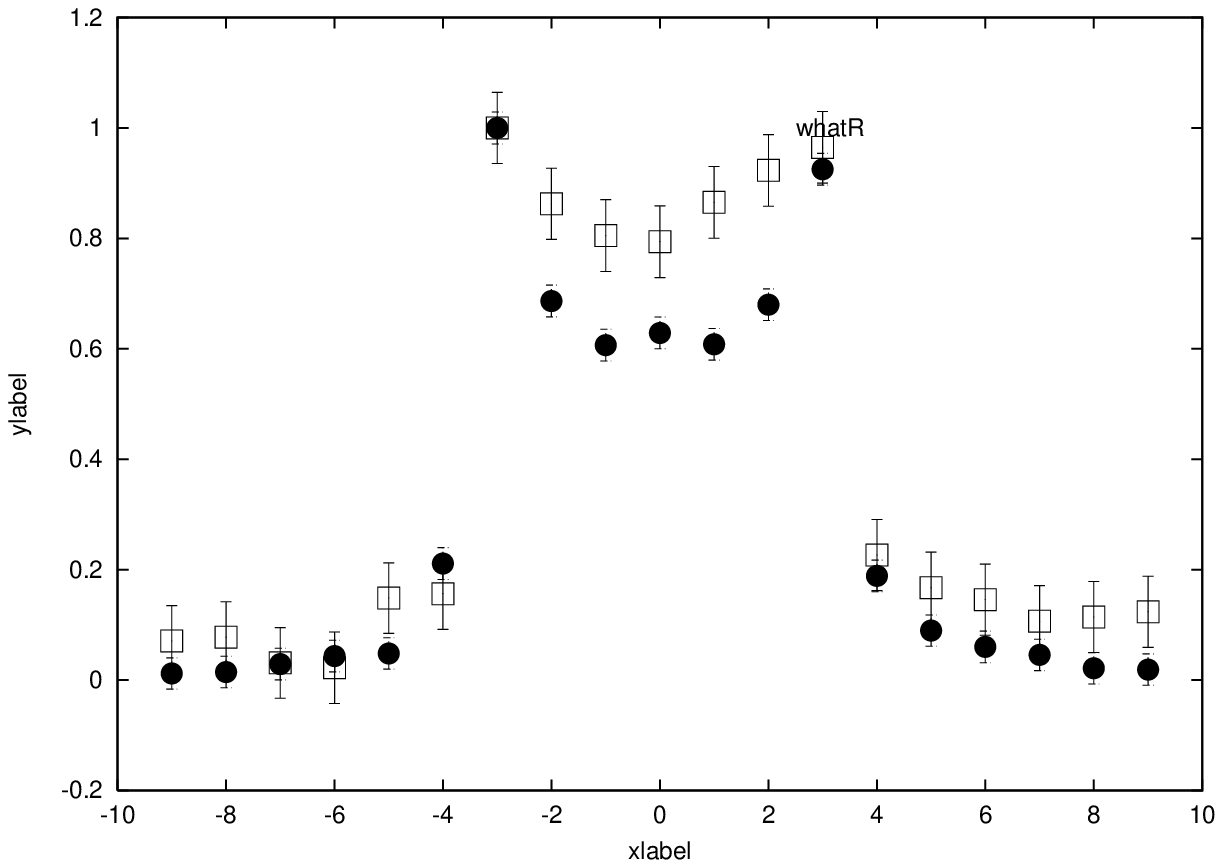}
\end{minipage}\\[1pc]
\begin{minipage}[b]{8.5pc}
\psfrag{whatR}[lb][lb]{\tiny $R/a_{s}=5$}
\psfrag{xlabel}[ct][ct]{\tiny $x_{\perp}/a_{s}$}
\psfrag{ylabel}[cb][cb]{\tiny $E^{\parallel}/E_{max}^{\parallel}$}
\includegraphics[width=8pc]{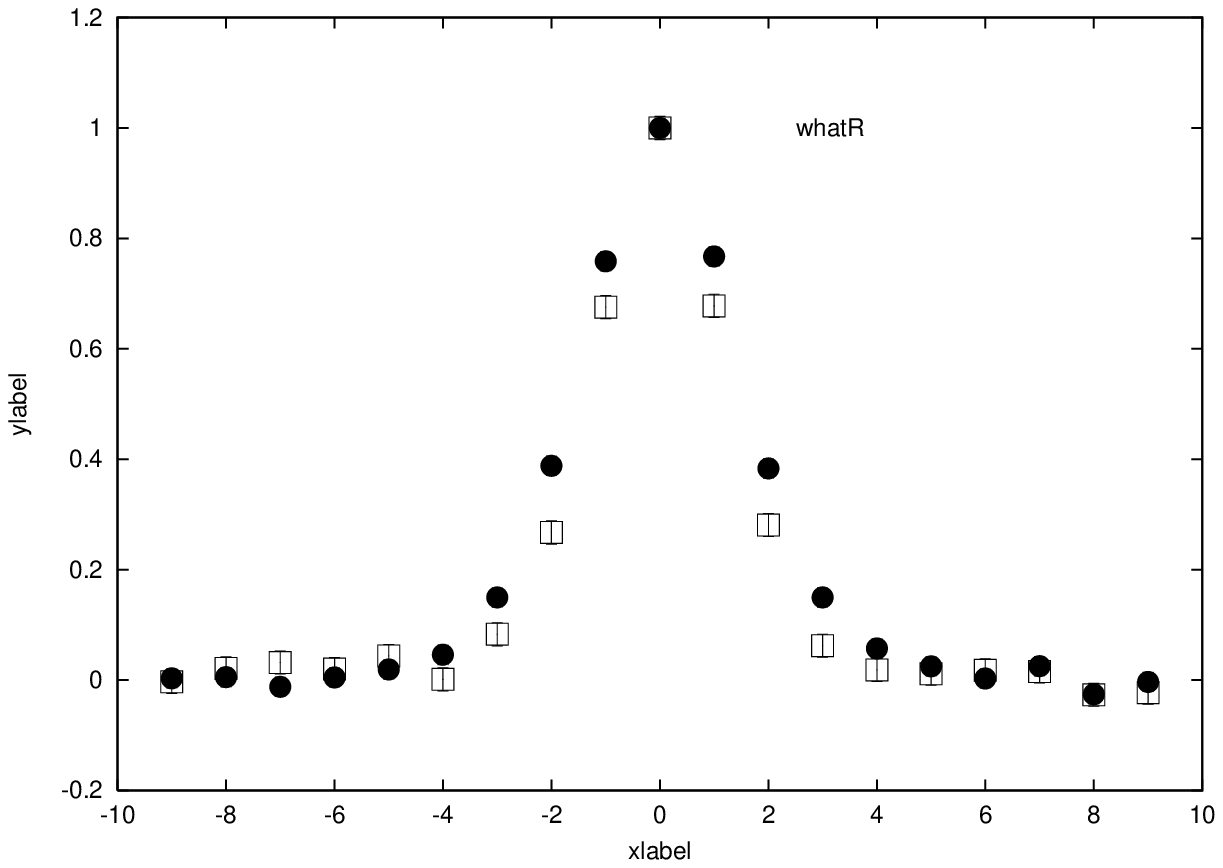}
\end{minipage}%
\begin{minipage}[b]{8.5pc}
\psfrag{whatR}[lb][lb]{\tiny $R/a_{s}=7$}
\psfrag{xlabel}[ct][ct]{\tiny $x_{\perp}/a_{s}$}
\psfrag{ylabel}[cb][cb]{\tiny $E^{\parallel}/E_{max}^{\parallel}$}
\includegraphics[width=8pc]{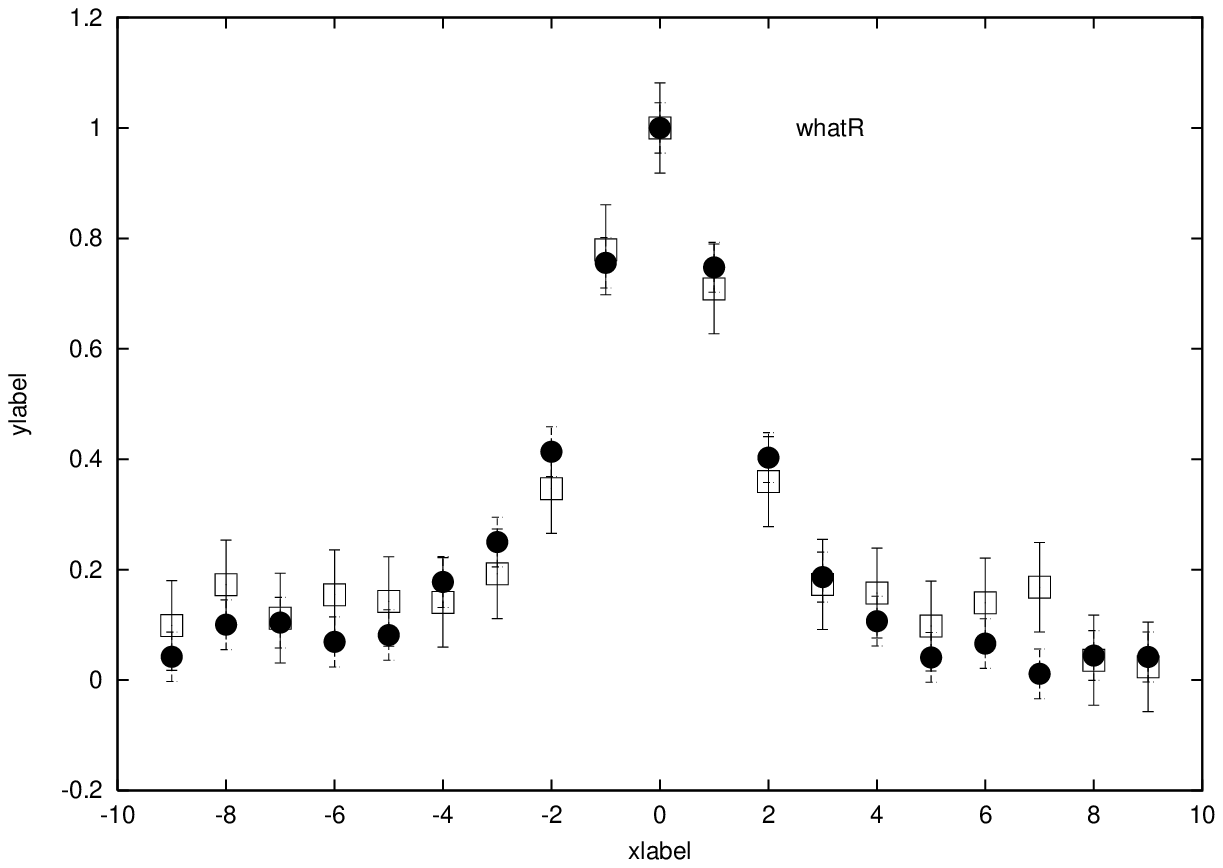}
\end{minipage}
\caption{Flux tube profiles for $R/a_{s}=5,7$. The open squares and
the solid circles are the quenched and unquenched results respectively.}
\label{fluxprofile}
\end{figure}

The lattice observable needed to measure the field strength
is given by the following correlator of plaquette
$P_{\mu\nu}$ with a Wilson loop $W(R,T)$ \cite{Trot93}
\begin{equation}
f_{\mu\nu}^{R,T}(\vec{x}) \sim
\frac{<W(R,T)
\left(P_{\mu\nu}(\vec{x})-P_{\mu\nu}(\vec{x}_{\infty})\right)>}
{<W(R,T)>}.
\end{equation}
Here, $\vec{x}$ is measured relative to the center of the Wilson loop and
$\vec{x}_{\infty}$ is usually taken as the site half the lattice away on
a finite size lattice. We measured the $f_{13}$ component, which is
corresponding to the energy density of the color-electric field in the
direction parallel to $R$ in the continuum limit,
i.e., $f_{13}=\mathcal{E}^{\parallel}$.
The flux tube profiles for quark separations $R/a_{s}=5,7$
are given in Fig. \ref{fluxprofile} where
$\mathcal{E}^{\parallel}(x)$ is plotted both along
($x_{\parallel}$) and perpendicular ($x_{\perp}$) to $R$.
The data is obtained with $T/a_{s}=7$.
Results here illustrate the formation of flux tubes between
the color charges. Notice that the thickness of the tube is
$\simeq 1.6fm$ and is independent of the quark separation $R$.

\begin{figure}[!t]
\psfrag{xlabel}[ct][ct]{$R/a_{s}$}
\psfrag{ylabel}[cb][cb]{$E^{\parallel}(0)_{unquenched}/
E^{\perp}(0)_{quenched}$}
\psfrag{T=2}[rb][rb]{\tiny $T/a_{s}=2$}
\psfrag{T=6}[rb][rb]{\tiny $T/a_{s}=6$}
\psfrag{T=8}[rb][rb]{\tiny $T/a_{s}=8$}
\psfrag{T=9}[rb][rb]{\tiny $T/a_{s}=9$}
\includegraphics[width=17.5pc]{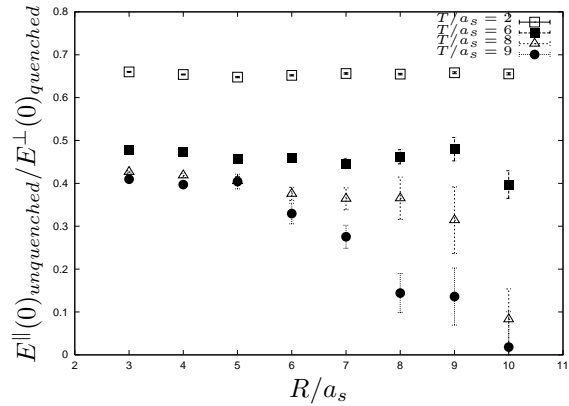}
\caption{Unquenched $E^{\parallel}(0)$ for different $R$.}
\label{fluxdrop}
\end{figure}

Fig. \ref{fluxprofile} also demonstrates the screening of the sources
by the dynamical quarks: the unquenched field strength is significantly
lower than the quenched results inside the flux tube. One would also
expect the unquenched field strength to drop substantially while the
quenched results stay fairly constant near the string breaking distance.
This idea is illustrated in Fig. \ref{fluxdrop} where the field strength at
the center of the tube is plotted against the quark separation for the
unquenched data. One can observe that at sufficiently large $T$,
the field strength decreases gradually when $R$ increases, and eventually
vanishes beyond the string breaking distance of $R/a_{s} \simeq 8.0-10.0$.

To summarize, we were able to observe string breaking by dynamical quarks in
$\mathrm{QCD_{3}}$. The use of improved actions on coarse lattices provided
a crucial advantage in accessing the quark separations and propagation
times at which string breaking occurs. The results confirm that string
breaking appears in Wilson loop correlators only at sufficiently large
$T$, of about $1fm$. These have clear implications
for simulations of full QCD in 4-d as this scale is
computationally accessible particularly on modestly coarse lattices. 

HDT is supported by NSERC.

\end{document}